\documentclass[runningheads]{llncs}
\usepackage{comment}
\usepackage{url}

\usepackage{graphicx}
\usepackage{float}
\usepackage{subcaption} 
\usepackage{xcolor}
\usepackage{tikz}
\usepackage{hyperref}
\usepackage{siunitx}

\begin{document}
\mainmatter              
\title{High-Performance Ptychographic Reconstruction 
with Federated Facilities}
\titlerunning{Federated Ptychographic Reconstruction}  
%
\author{Tekin Bicer\inst{1,3} \and Xiaodong Yu\inst{1} \and
Daniel J. Ching\inst{3} \and Ryan Chard\inst{1} \\ Mathew J. Cherukara\inst{3} \and Bogdan Nicolae~\inst{2} \and Rajkumar Kettimuthu\inst{1} \and Ian T. Foster\inst{1}}
\authorrunning{Tekin Bicer et al.} 
%
\tocauthor{Tekin Bicer, Xiaodong Yu, Daniel Ching, Ryan Chard, Mathew J. Cherukara, Bogdan Nicolae,
Rajkumar Kettimuthu, and Ian T. Foster}
\institute{
Data Science and Learning Division, CELS\\
\and
Mathematics and Computer Science Division, CELS\\
\and
X-ray Science Division, APS\\
Argonne National Laboratory, Lemont IL 60439, USA\\
\email{\{tbicer,xyu,dching,rchard,mcherukara,bnicolae,kettimut,foster\}@anl.gov}\\
}

\maketitle              

\begin{abstract}

Beamlines at synchrotron light source facilities are 
powerful scientific instruments used to image samples and observe phenomena at high spatial and temporal resolutions. Typically, these facilities are equipped only with modest compute resources for the analysis of generated experimental datasets. However, high data rate experiments can easily generate data in volumes that take days (or even weeks) to process on those local resources. 
To address this challenge, we present a system that unifies leadership computing and experimental facilities by enabling the automated establishment of data analysis pipelines that extend from edge data acquisition systems at synchrotron beamlines to remote computing facilities; under the covers, our system uses Globus Auth authentication to minimize user interaction, funcX to run user-defined functions on supercomputers, and Globus Flows to define and execute workflows. We describe the application of this system to ptychography, an ultra-high-resolution coherent diffraction imaging technique that can produce 100s of gigabytes to terabytes in a single experiment.
When deployed on the DGX A100 ThetaGPU cluster at the Argonne Leadership Computing Facility and a microscopy beamline at the Advanced Photon Source, our system performs analysis as an experiment progresses to provide timely feedback.

\end{abstract}

\section{Introduction}
Synchrotron light sources are used by thousands of scientists from a wide variety of communities, such as energy, materials, health, and life sciences, to address challenging research problems~\cite{aps-science-2020,aps-web} by providing unique tools for materials characterization. A subset of these tools includes coherent imaging techniques which enable \textit{in-situ} and \textit{operando} studies of functional, structural, and energy materials at high-spatial resolution.

\begin{figure}[t]
\centering
\subcaptionbox{Reconstructed image from 2D diffraction data.}{
    \begin{tikzpicture}
        \node[anchor=south west,inner sep=0] at (0,0) {
            \includegraphics[width=0.30\textwidth]{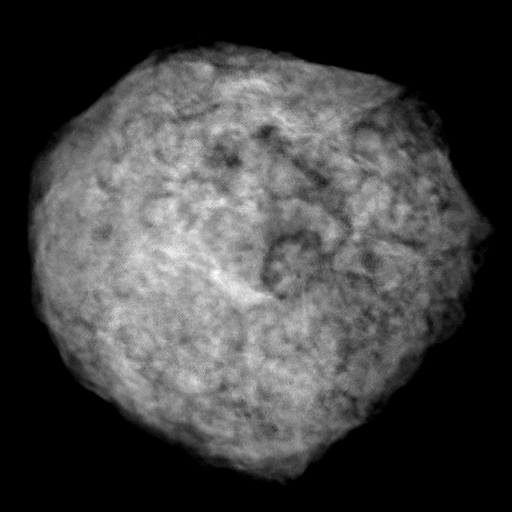}};
        \draw[red,ultra thick] (1.5,1.6) circle (0.07);
    \end{tikzpicture}
}%
\hfill 
\subcaptionbox{2D cropped diffraction pattern.}{
    \includegraphics[width=0.30\textwidth]{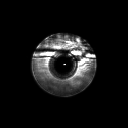}
}%
\hfill 
\subcaptionbox{Visualization of pixels with log(b)}{
    \includegraphics[width=0.30\textwidth]{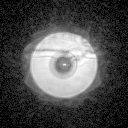}
}%
\caption{(a) shows a sample ptychographic reconstruction of a catalyst particle view. The red circle shows the location of an X-ray measurement. The corresponding (cropped) diffraction pattern of this measurement is shown in (b). During a ptychography experiment, many of these diffraction patterns are collected. (c) shows the same diffraction pattern after taking its log, highlighting the distribution of the pixel values. The outer values of the diffraction pattern carry information about the sharpness/corners of the sample features.}
\label{diff-img}
\end{figure}

A nanoscale imaging technique of increasing importance to both x-ray and electron microscopes is ptychography~\cite{pfeiffer2018x,Chen826,jesse2016big}. This non-invasive 2D imaging technique is widely used at synchrotron light sources to study functional, structural, biological, and energy materials at extremely high spatial resolutions.
During a ptychography experiment, a sample is continuously raster-scanned using a focused X-ray beam and the corresponding diffraction patterns are acquired on a photon-counting pixelated detector. These diffraction patterns are then processed using an iterative ptychographic reconstruction method to generate 2D real-space projection images (\autoref{diff-img}.) Although ptychography involves high photon cost, it can deliver extremely high spatial resolutions, enabling imaging of (bio)samples, for example, trace elements of green algae at sub-30-nm~\cite{deng2015simultaneous}, bacteria at 20-nm~\cite{wilke2012hard} and diatoms at 30-nm resolution~\cite{vine2012simultaneous}. 
Ptychography is already used at many synchrotron light source beamlines, including Advanced Photon Source (APS) and National Synchroton Light Source II (NSLS-II), and is expected to be yet more common at next generation light sources~\cite{APSEarlySience2015,apsu} where the required photon budget is easier to meet.

Ptychography experiments can generate data at high rates over extended periods.
For example, detectors currently used in ptychographic experiments at synchrotron light sources can generate 1030$\times$514 12-bit pixel frames at 3~kHz, yielding a 19.5~Gbps data generation rate. Next-generation light sources, such as the APS upgrade (APSU)~\cite{apsu}, are expected to increase X-ray beam brightness by more than two orders of magnitude, an increase that will enable lensless imaging techniques such as ptychography to acquire 
data at MHz rates~\cite{hammer2021strategies}, potentially increasing data acquisition rates to Tbps. Such dramatically greater data acquisition rates are scientifically exciting but also pose severe technical challenges for data processing systems. It is expected that a single ptychography experiment will soon be able to  generate 
many PBs of raw and reconstructed data, pushing the limits of I/O and storage resources even for high-performance computing resources and superfacilities~\cite{huang2021fast,enders:2020:superfacility,wang2021deploying,bicer2016optimization}.

Greatly increased data rates and volumes also pose major challenges for the reconstruction computations used to recover real-space images from the diffraction pattern data obtained via ptychographic imaging. 
The ptychographic reconstruction process is typically data-intensive, requiring hundreds of iterations over diffraction patterns and the reconstructed object.
Moreover, if the goal is to recover a 3D volumetric image, then tomographic (or laminographic) reconstruction techniques need to be performed after ptychographic reconstruction~\cite{hidayetouglu2019memxct,venkatakrishnan2021algorithm},  further increasing the computational demand and execution time of the processing pipeline~\cite{aslan2019joint,nikitin2019photon}.

Today's state-of-the-art ptychographic data analysis workflows mostly utilize locally available compute resources, such as high-end beamline workstations, or small clusters due to the difficulties in  accessing remote HPC resources and/or scaling algorithms on large-scale systems. Further, the workflows are typically executed only after the data acquisition is finalized.
This type of offline data analysis is not feasible for experiments that generate massive measurement data, and will generally be impossible to perform with the next generation light sources. 
Increasingly, therefore, the need arises to run data analysis workflows on specialized high-performance (HPC) systems in such a way that data can be analyzed \emph{while an experiment is running}.
However, effective federation of instrument and HPC system introduces many technical problems, from user authentication to job scheduling and resource allocation, transparent data movement, workflow monitoring, and fault detection and recovery.
Robust solutions to these problems require sophisticated methods that for widespread use need to be incorporated into 
advanced software systems.

We present here a system that we have developed to implement solutions to these problems. 
This system unifies HPC and experimental facilities to enable on-demand analysis of data from a ptychographic experiment while the experiment is running. Its implementation leverages a suite of cloud-hosted science services provided by the Globus platform:   Globus Auth for authentication, so as to avoid the need for repeated user authentication~\cite{tuecke2016globus}; Globus transfer for rapid and reliable data movement between light source and HPC~\cite{chard2016globus}; funcX for remote execution of user-defined functions on remote compute resources~\cite{chard2020funcx}; and Globus Flows to coordinate the multiple actions involved in collecting and analyzing data~\cite{ananthakrishnan2018globus}.

The rest of this paper is organized as follows. In \autoref{sec:background}, we briefly explain data acquisition and analysis steps for ptychography. In \autoref{sec:engine}, we present components of our system and their interaction. We evaluate our system and its end-to-end performance in \autoref{sec:experiment}, discuss related work in \autoref{sec:related}, and conclude in \autoref{sec:conclusion}.
\section{Background}
\label{sec:background}

\begin{figure}[t]
    \centering
    \includegraphics[width=1\textwidth]{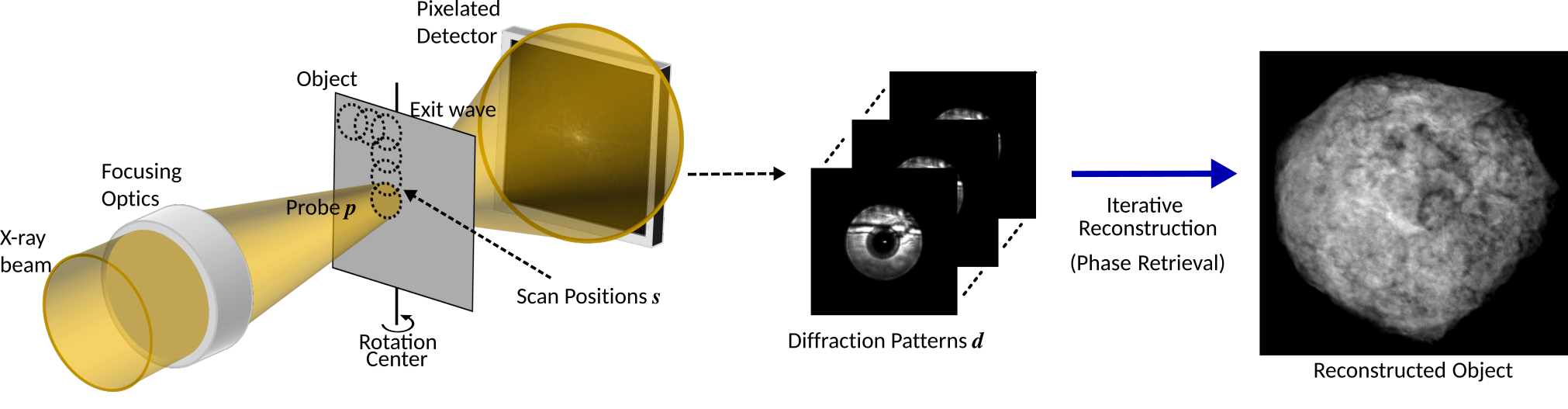}
    \caption{Ptychography experiments and ptychographic reconstruction}
    \label{fig:ptyc}
\end{figure}

\autoref{fig:ptyc} illustrates the experimental setup for ptychography and the basic steps for the 2D reconstruction. During a ptychography experiment, an object is placed on top of a rotation stage and scanned by a focused X-ray beam (probe $p$). As the object is being illuminated by an X-ray beam, the far-field diffraction patterns ($d$) are collected from a series of overlapping positions ($s$) using a pixelated detector. This allows large objects to be scanned and imaged with a higher resolution than the beam size.

At the end of a ptychography experiment, a 3D experimental dataset, which consists of 2D diffraction pattern images from a fixed rotation, is generated. The size of the dataset depends on many factors, such as target resolution, size of the object, overlapping area, and size of the detector. The data acquisition rates of  ptychography experiments are typically proportional to the beam intensity. For example, while lab systems require significant scanning time over a small area to meet the photon requirements~\cite{batey2021x} (e.g., ``16 hours for \SI{400}{\micro\meter}$^2$ area at 400 fps''), synchrotron light sources can provide bright beams and enable imaging cm$^2$ area at several kHz (and MHz in the future) with much higher resolutions.

Ptychographic reconstruction takes a set of diffraction patterns, $d$, with their corresponding scanning positions, probe information $p$, and the initial guess of the object ($\psi^{i}$), and then tries to iteratively converge an object, $\psi^{i+1}$, that is consistent with the measurement data while solving the phase retrieval problem as shown in \autoref{eq:ptych}. This process typically requires hundreds of iterations on a large dataset, therefore it is an extremely data-intensive process. Most state-of-the-art implementations rely on accelerators such as GPUs. 

\begin{equation}
    \label{eq:ptych}
    \psi^{i+1} = F(d, p, \psi^{i})
\end{equation}

Ptychographic reconstruction aims to recover a single 2D real-space projection using a set of diffraction patterns that are collected from a fixed rotation angle. This data acquisition scheme can be repeated for different angles, adding angle dimension to $d$, which, in turn, extends the 2D ptychography imaging technique to 3D ptychography (or ptychographic tomography)~\cite{holler2014x}. This compound technique can image 3D volumes of samples at extremely high spatial resolutions, for example, integrated circuits at sub-10-nm~\cite{aps-raven} and nanoporous glass at 16-nm~\cite{holler2014x}. However, the new ptychography dataset is typically more than two orders of magnitude larger compared to the single 2D ptychography dataset ($d$), which results in a significant increase in both storage and compute requirements of the analysis tasks. 

Ptychographic tomography problem can be solved with several approaches, including the {\em two-step} and more advanced {\em joint} approaches~\cite{aslan2019joint,chang2019blind}. The two-step approach treats ptychography and tomography as sequential problems, first recovering a 2D real-space projection for each rotation angle and then performing tomographic reconstruction on all projections to generate a 3D object volume. Joint approaches, on the other hand, consider the ptychography and tomography problems as one problem and continuously use information from both ptychography and tomography during reconstruction. The tomography problem can be solved by using high-performance advanced reconstruction techniques~\cite{venkatakrishnan2016robust,chantzialexiou2018pilot}; however, in both two-step and joint approaches, the added 3D reconstruction operations translate to additional compute resource requirements for the workflow~\cite{nikitin2019photon}.
\section{Ptychography Workflow with Federated Resources}
\label{sec:engine}

We now introduce our system and the ptychography workflow that is executed. Our system aims to automate workflow execution on geographically distributed facilities and resources using Globus services and performs high-performance ptychographic reconstruction. We describe the main components of our workflow system and how they interact with each other in the following subsections.

\subsection{Automated Light Source Workflow Execution and Coordination}
\begin{figure}[t]
    \centering
    \includegraphics[width=1\textwidth]{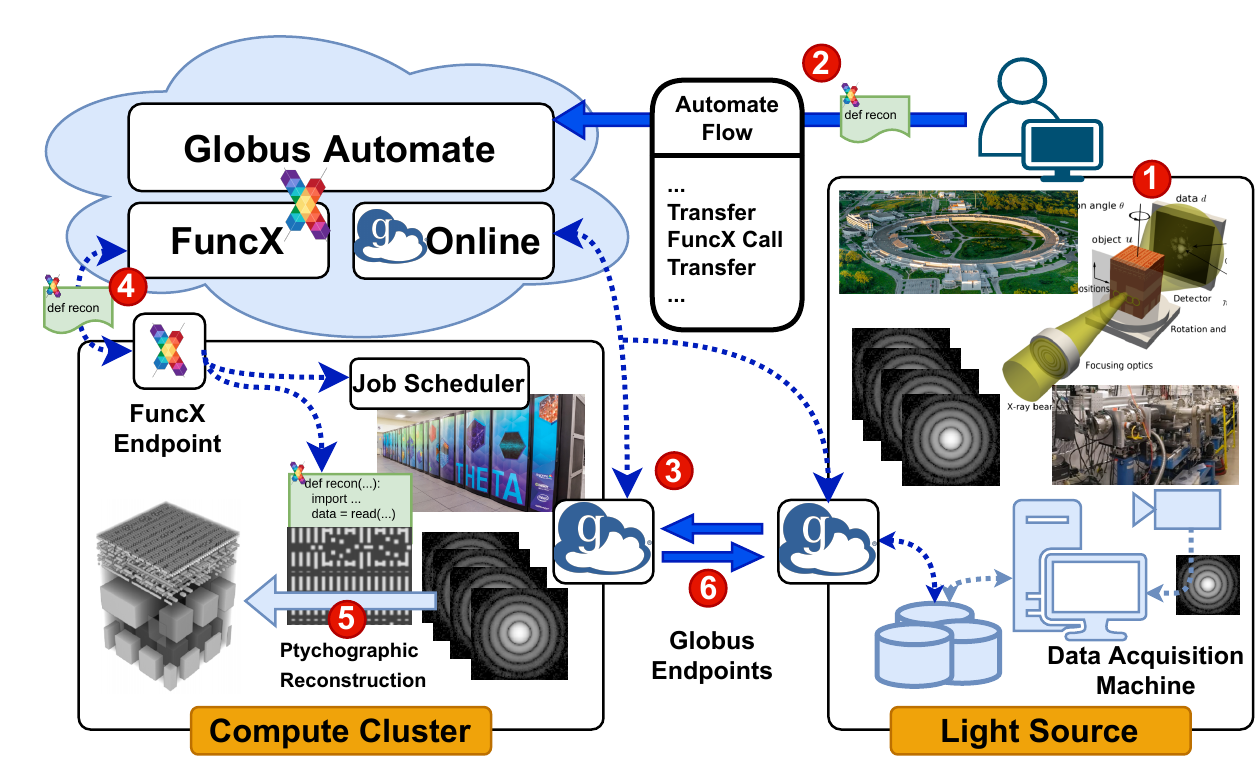}
    \caption{System components and execution flow}
    \label{fig:system}
\end{figure}

The ptychography workflows start with the data acquisition step as mentioned in the previous sections.
The diffraction pattern data is collected at the detector and continuously streamed to the data acquisition machine. This process is illustrated with step (1) in \autoref{fig:system}.

We use {\em Globus Flows} to describe and execute ptychography workflows~\cite{ananthakrishnan2018globus}. This cloud-based service is 
designed to automate various data management tasks such as data transfer, analysis, and indexing. 
Workflows in Flows are defined with a JSON-based state machine language, which links together calls to external \textit{actions}. 
Flows implements an extensible model via which external \textit{action providers}
can be integrated by implementing the Globus Flows action provider API. 
At present, Flows supports ten actions (e.g., transfer data with Globus Transfer, execute functions with funcX, or associate identifiers via DataCite) from which users can 
construct workflows. 
Flows relies on Globus Auth~\cite{tuecke2016globus} to provide secure, authorized, and delegatable access to user-defined workflows and also the Globus-auth secured action providers.

We specify the ptychography workflows as a Flows flow definition. This flow consists of three main {\em actions}: (i) {\em transfer} data from data acquisition machine (edge) at light source to compute cluster, e.g., ThetaGPU at Argonne Leadership Computing Facility; (ii) initiate {\em reconstruction process} via remote function call; and (iii) {\em transfer} reconstructed images from compute cluster back to light source. This flow definition is submitted to the Flows cloud service as shown in step (2) in \autoref{fig:system}. Once the flow definition is deployed to the Flows service, any number of flows can be initiated with workflow-specific input parameters. 

Recall that ptychographic reconstruction process takes 2D diffraction patterns collected from a specific angle/view and recovers the corresponding 2D projection image. 
For 3D ptychography, the sample is rotated many times. The diffraction patterns collected from each angle can be reconstructed independently. 
Thus, we can write a single (generic) flow definition and reuse it with different reconstruction parameters for different angles. Therefore, a 3D ptychography workflow typically consists of many independent sub-workflows, each executing the same flow definition. 
A sample ptychography workflow definition can be found at DOE-funded Braid project~\cite{www:braid}.

\subsection{Transparent Remote Function Calls and Data Transfers}
The ptychography flow uses Globus transfer for efficient cross-facility data transfers.
We deployed a Globus transfer endpoint to the beamline data acquisition machine. 
During the ptychography experiment, each scan is saved into a folder with a unique id, e.g., \texttt{scan100} or \texttt{flyscan100}. Our system reads these folder names, extracts the unique ids, and generates corresponding input and output folders at the beamline and the compute resource endpoints.
For example, if ptychography data is stored in  \texttt{$<$prefix$>$/input/scan100} at a beamline endpoint, our system will create  \texttt{$<$prefix$>$/input/100} and \texttt{$<$prefix$>$/recon/100} 
directories at compute resource endpoint for transferring measurement data and saving (intermediate) reconstructed images. 

After relevant folders are created and ptychography data is transferred to the compute endpoint, the ptychographic reconstruction tasks are initiated at the compute cluster.
We use funcX for executing ptychographic reconstruction operations on remote computing clusters.
funcX is a function as a service (FaaS) platform for secure and distributed function execution while supporting dynamic provisioning and resource management. 
funcX uses Globus Auth for secure access to remote computing systems and thus interoperates seamlessly with Globus Flows. 
In a similar manner to Globus transfer, funcX relies on endpoints deployed on remote computing systems to enable function execution. 

We implemented the ptychographic reconstruction task as a Python function and registered it with the funcX cloud service. This operation serializes the reconstruction code, stores it in the cloud-hosted funcX service, and returns a unique identifier which can then be used to invoke the function.
The reconstruction function can be invoked on-demand on any accessible funcX endpoints, providing flexibility for running any funcX function on any active endpoint the user has permission to use.
If there are insufficient resources for the reconstruction task, the funcX endpoint dynamically requests additional resources (e.g., via Cobalt job scheduler at the ThetaGPU, ALCF). funcX automatically deploys {\em worker daemons} to newly allocated resources. These workers receive serialized reconstruction tasks and execute them.

Our ptychographic reconstruction code uses GPUs to perform analysis. If more than one reconstruction task is executed on a compute node, for example when a node has more than one GPU and each task requires only a single GPU, then each task needs to be pinned to a different GPU to maximize resource utilization.
funcX functions can scale efficiently on CPU-based compute resources since operating systems handle load balancing. However, accelerator-based compute resources, e.g., GPUs, are typically managed by device drivers such as CUDA. One way to perform load balancing between tasks is to use inter-process communication, but this is nontrivial for stateless (lambda-like) funcX functions. We implemented a file-based synchronization mechanism on shared memory \texttt{tmpfs} to track available GPUs on allocated compute nodes. Specifically, when a worker starts executing a reconstruction task, it first tries to acquire an exclusive lock using \texttt{fcntl} on a predefined file (e.g., \texttt{/dev/shm/availgpus}), that keeps track of the GPUs. Once the lock is acquired, the reconstruction task checks the available GPUs from the file and updates them (setting a set of GPUs busy) according to its resource requirements. Since the number of GPUs and workers are limited on a compute node, the contention on the file is minimal and the performance bottleneck due to the (un)lock operations is negligible. Step (4) in \autoref{fig:system} shows the interaction between compute cluster and the funcX service.

\subsection{Accelerated Ptychographic Image Reconstruction}
Ptychographic reconstruction is an iterative process that can be extremely data-intensive, depending on the dataset and reconstruction method. Efficient reconstruction of ptychography data and timely feedback are crucial for relevant data acquisition, early detection of experimental errors, and steering experiments.

Our ptychographic reconstruction workflows rely on our in-house developed parallel ptychographic reconstruction code~\cite{tike,yu2021scalable}. Specifically, we implemented several parallel solvers, including multi-GPU conjugate-gradient and least-squares gradient descent solvers, to use in reconstructions. Our advanced parallelization methods provide efficient topology-aware communication between reconstruction threads, while mapping communication (synchronization) intensive threads to GPUs connected with high-performance interconnects, such as NVlink pairs and switch~\cite{yu2021ptycho}. These parallelization techniques enable us to efficiently scale a single reconstruction task to multiple GPUs. We evaluate the scalability performance  in the following section.

After the reconstruction task is completed, the funcX endpoint informs the Flows service, which then executes the next state, and initiates another Globus transfer operation to retrieve reconstructed images from compute cluster to beamline at synchrotron light source. The reconstruction and final data transfer steps are shown in steps (5) and (6), respectively.
\section{Experimental Results}
\label{sec:experiment}

We evaluated the performance of our ptychographic reconstruction workflow in a configuration that connected  the 2-ID microscopy beamline at the APS synchrotron light source facility and the ALCF HPC facility,
located $\sim$1 km from APS at Argonne National Laboratory.
To permit detailed and repeated evaluations, we did not perform actual ptychographic experiments in these studies but instead ran a program on the 2-ID data acquisition computer that replayed images at an appropriate rate.
The data acquisition machine has a $\sim$30 Gbps connection to the detector and a 1 Gb Ethernet connection to outside.

We used four datasets to evaluate our system: a real-world catalyst particle dataset, and three phantom datasets: two coins and a Siemens star, respectively.
The datasets have different dimensions and thus different computational requirements. 
Specifically, the catalyst particle dataset is a 3D ptychography dataset of 168 views/angles, each with $\sim$1.8K diffraction patterns with dimensions 128$\times$128, for a total size of 168$\times$1800$\times$128$\times$128$\times$4B = 20~GB. 
The coin and Siemens star datasets are 2D ptychography datasets with 8K, 16K, and 32K diffraction patterns, respectively, all of size (256, 256), 
for total dataset sizes of 2~GB, 4~GB, and 8~GB, respectively.

The datasets are reconstructed on the ThetaGPU cluster at ALCF, which consists of 24 DGX A100 nodes, each with eight NVIDIA A100 accelerators connected with NVSwitch. Each A100 GPU has 40~GB memory. The host machine 
has two AMD Rome CPUs and 1~TB DDR4 memory. The ThetaGPU nodes are connected to Grand, a 200~PB high-performance parallel file system. Since allocation of compute nodes from shared resources (using job scheduler) can introduce significant overhead, we reserved the compute nodes in advance. Our reconstruction jobs still use the job scheduler, however the queue wait time is minimized due to reserved nodes.

The reconstructions are performed by using the Tike library~\cite{tike}, which provides parallel ptychographic reconstruction capabilities in multi-GPU settings \cite{yu2021ptycho,yu2021scalable}. We used a conjugate gradient solver and reconstructed both object and probe.
The object is partitioned in grid cells and the neighboring cells are synchronized at the end of each iteration.

\subsection{Optimum GPU Configuration}
\label{sec:opt-gpu}
\begin{figure}[t]
\centering
\subcaptionbox{Catalyst particle\label{single-catalyst}}
    {\includegraphics[height=0.375\textwidth]{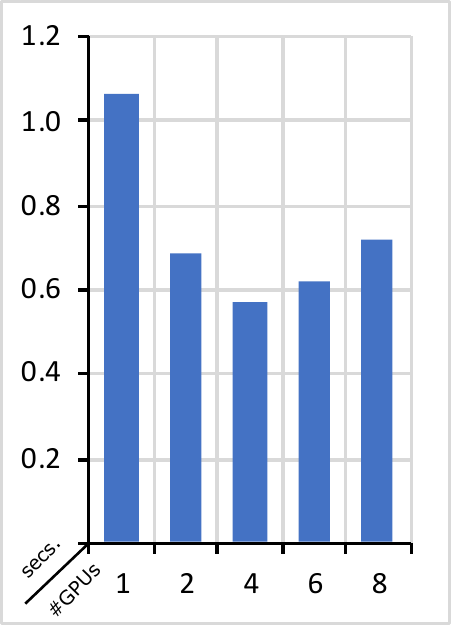}}%
\hfill
\subcaptionbox{Coin 8K\label{single-coin8k}}
    {\includegraphics[height=0.375\textwidth]{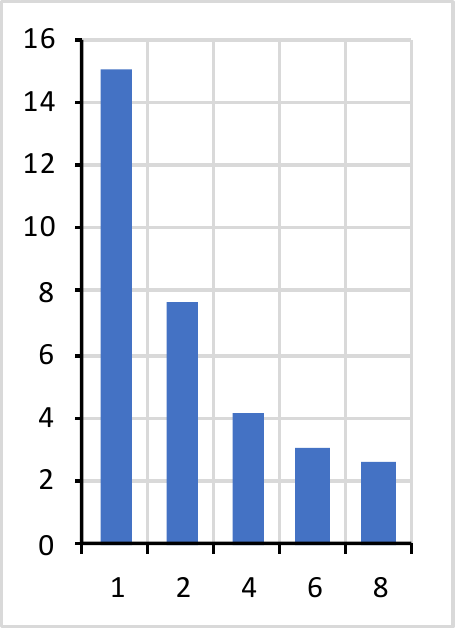}}%
\hfill
\subcaptionbox{Coin 16K\label{single-coin16k}}
    {\includegraphics[height=0.375\textwidth]{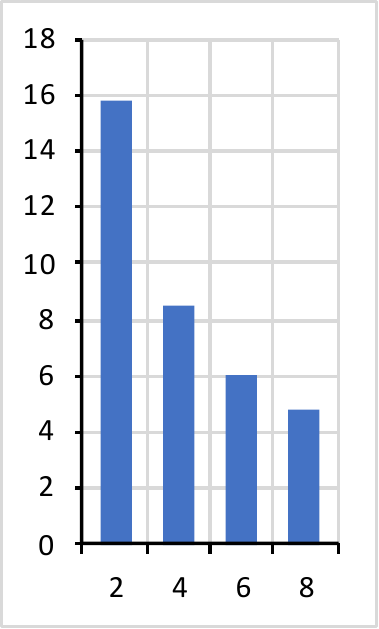}}%
\hfill
\subcaptionbox{Siemens star\label{single-siemens}}{\includegraphics[height=0.375\textwidth]{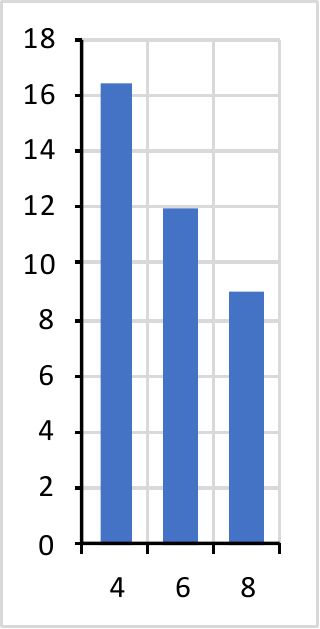}}%
\caption{Reconstruction times for different ptychography datasets on an eight-GPU DGX A100 node on the ThetaGPU cluster. The x-axis shows the number of GPUs used for reconstruction, and the y-axis shows the per iteration execution time in seconds. We reconstructed only a single view/angle from each dataset. The dimensions of datasets are (1.8K, 128, 128), (8K, 256, 256), (16K, 256, 256), and (32K, 256, 256) for catalyst particle, coin 8K, coin 16K, and Siemens star, respectively. Catalyst particle is a real experimental dataset collected at APS, whereas the others are synthetically generated.}
\label{fig:scale}
\end{figure}

We first conduct experiments to determine the optimum GPU configuration for the reconstruction computation. We perform a 50-iteration reconstruction for each of our four datasets on each of $\{1, 2, 4, 6, 8\}$ GPUs.
We present the {\em per iteration} reconstruction time with respect to corresponding GPU configuration in \autoref{fig:scale}.

Reconstruction of catalyst particle, as shown in \autoref{single-catalyst}, can (sub-optimally) scale 
up to four GPUs, then the inter-GPU communication becomes the bottleneck and starts introducing overhead. The speedups for the two and four-GPU configurations relative to the one-GPU configuration are only 1.56 and 1.88, respectively. The catalyst particle is a small dataset, where each view is $\sim$113MB, therefore its computational demand is minimal ($\sim$1 second per iteration). This results in sub-optimal scaling efficiency and favors single-GPU reconstruction.

\autoref{single-coin8k} shows scaling results for the synthetic coin dataset (with 8K diffraction pattern). We observe good scaling efficiency on up to four GPUs, achieving speedups of 1.96 and 3.6 on two and four GPUs, respectively. On more GPUs, we see diminishing returns due to communication and observe lower scaling efficiencies, ranging from 72--82\%.

\autoref{single-coin16k} shows the same synthetic coin dataset with a larger number of diffraction patterns. The memory footprint of this dataset is significantly larger than with the previous datasets and cannot fit on one GPU; hence the missing configuration. The larger dataset translates to more computational load and therefore we see improved GPU scaling performance: more than 90\% scaling efficiency for the four-GPU configuration and more than 80\% for the rest. 

Lastly, \autoref{single-siemens} shows the per iteration reconstruction times with the largest dataset. Similar to the previous dataset, this dataset has a large memory footprint and can be reconstructed only with more than four GPUs. Since the dataset is large enough, its scaling efficiencies are larger than 90\% for six and eight GPUs when compared to the four-GPU configuration. 

When we perform a cross-comparison between 8K and 16K versions of the coin datasets, and the Siemens star dataset (32K diffraction patterns), we see a good weak scaling performance for the reconstruction tasks. Specifically, comparing the two-GPU configuration of 8K and four-GPU configuration of 16K coin datasets, and the eight-GPU configuration of 32K Siemens dataset, shows $>$90.5\% weak scaling efficiency. 

\subsection{End-to-end Workflow Evaluation}
\begin{figure}[t]
\centering
\subcaptionbox{Single node workflows\label{wf-all}}{
\centering\vspace{-9mm}\includegraphics[height=0.52\textwidth]{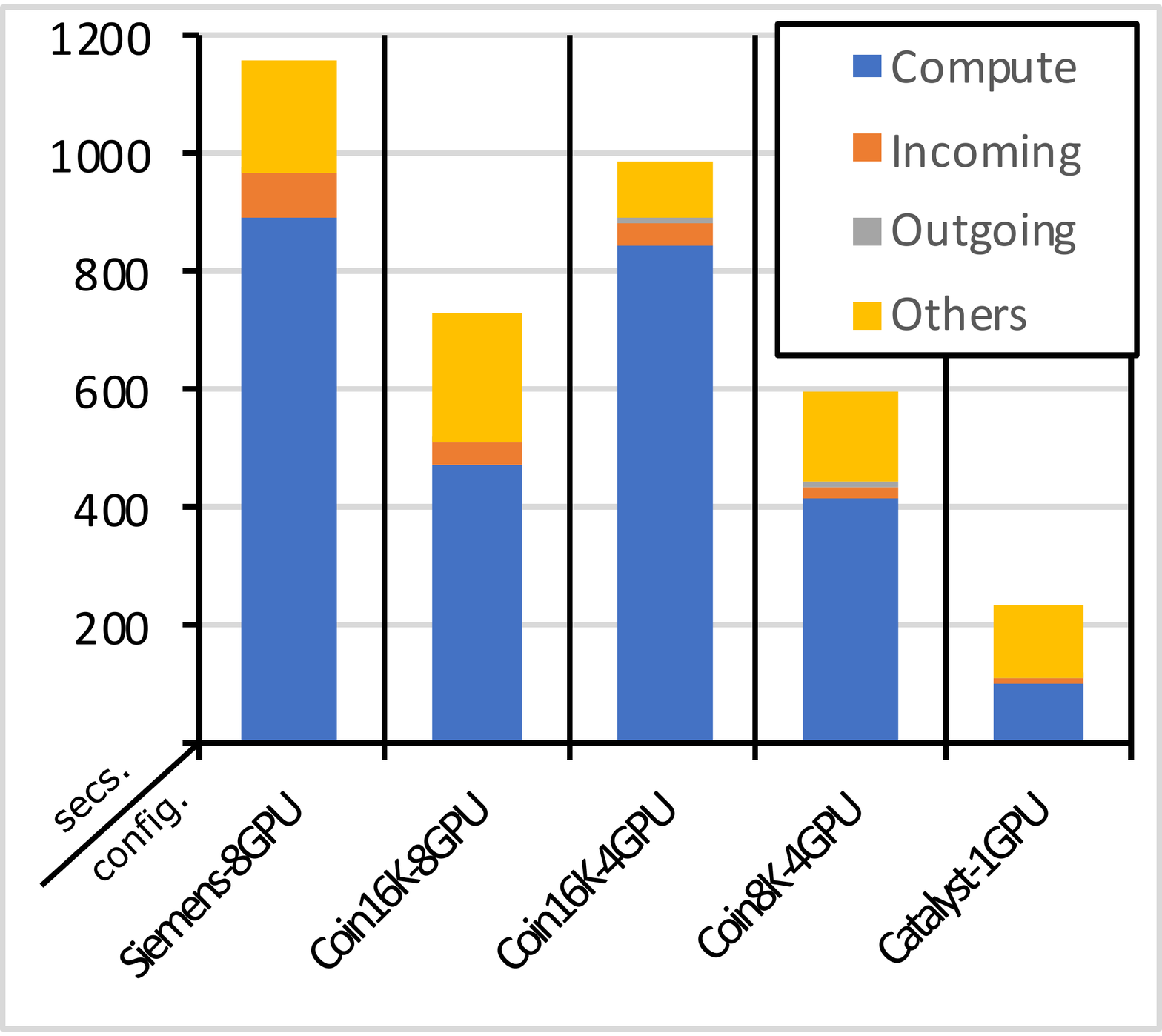}}%
\hfill
\centering
\subcaptionbox{Multi-node catalyst particle workflow\label{wf-catalyst}}{
   \centering\vspace{-17mm}\includegraphics[height=0.65\textwidth]{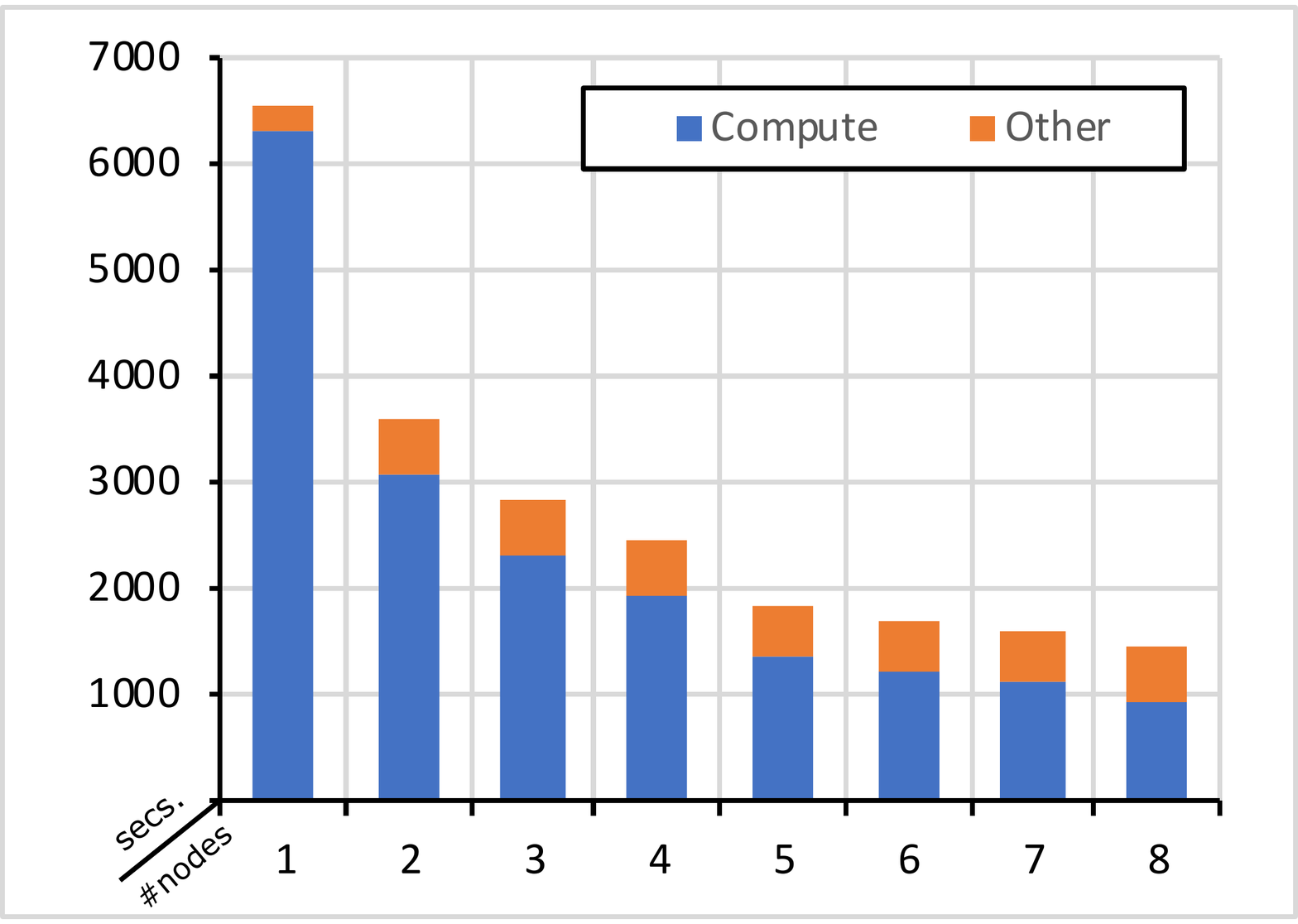}
}%
\caption{End-to-end performance of ptychographic reconstruction workflows. (a) shows the performance (y-axis, in seconds) for different datasets (x-axis) on a single DGX A100 node (up to eight GPUs.) The performance information is broken down according to the stages of workflows. (b)  shows the execution times for the 3D catalyst particle dataset. This workflow consists of 168 independent sub-workflows that can be executed concurrently. The x-axis shows the number of nodes (up to eight nodes or 64 GPUs) and the y-axis shows the total execution time (in seconds) for reconstructing all sub-workflows.} 
\label{fig:scale2}
\end{figure}

In these experiments, we evaluate the end-to-end execution of ptychography workflows using our system. We configure the number of GPUs according to our single-node performance results in \autoref{sec:opt-gpu}.

Our initial experimental setup focuses on ptychographic reconstruction workflow for single (2D) view datasets on single-GPU. Our workflow consists of several (potentially overlapping) steps.  
Specifically, our workflow consists of the following stages: (i) a user-defined Globus Flows script is deployed to the Flows service; (ii) the Flows service initiates a Globus transfer from the beamline data acquisition machine to the ThetaGPU parallel file system; (iii) once the transfer is complete, the Flows service executes the next state, triggering a user-defined ptychographic reconstruction funcX function via the funcX service. At this point, if the funcX endpoint has insufficient compute resources, it interacts with the resource management system, i.e., the job scheduler, to allocate additional resources. Last, when the funcX function is finalized, (iv) the Flows service initiates another data transfer to return reconstructed images to the beamline data acquisition machine at APS.

\autoref{wf-all} shows the execution of this workflow. The \texttt{compute} column shows the reconstruction time, and \texttt{incoming} and \texttt{outgoing} columns represent the transfers between data acquisition machine and ThetaGPU. Finally, \texttt{others} shows the additional overheads, including initial resource allocation (job submission and queue wait time) and cloud service calls (interacting with Globus Flows, Transfer, and funcX services). Recall that we reserve the compute nodes in advance in order to minimize the overhead due to the queue wait time, however our system still requests compute resources from scheduler and this introduces a delay. The name of the configurations shows the dataset name and the number of GPUs used for reconstruction, e.g., Siemens-8GPU refers to the workflow of the Siemens star dataset where eight GPUs are used for reconstruction. We set the total number of iterations to 100 for these experiments.

Our first observation is the consistent reconstruction times among datasets. Specifically, we see more than 85\% weak scaling performance among Siemens, Coin8K and Coin16K datasets, which follow our results in the previous section. 
Our second observation is the overhead due to \texttt{other} operations in the workflows. Since the system interacts with many external (cloud) services and the job scheduler, there is a significant noise during the execution. The effect of this overhead drops as the dataset sizes increase. We observe 10--30\% overhead from \texttt{others} for Siemens, Coin16K and Coin8K datasets, and close to 50\% overhead for Catalyst (which is the computationally least demanding dataset.) Lastly, we see the highest data transfer time for the Siemens star since it is the largest dataset. The data transfer times take 2.9-6.7\% of the end-to-end execution times.

In \autoref{wf-catalyst}, we present the multi-node workflow performance results for the catalyst particle dataset. This dataset consists of 168 views, which translates to 168 independent workflows that can be executed concurrently. We use up to eight ThetaGPU nodes (or 64 total GPUs). Each view is reconstructed using a single GPU.

The concurrency between sub-workflows enables overlapping aforementioned workflow steps and therefore partially hides the overhead of the steps. Since each view is being reconstructed using one GPU, the maximum depth of the concurrency is 64 (from eight GPU nodes.) Although this concurrency provides better performance, it also complicates the breakdown of execution times. Therefore, we only show the \texttt{compute} and \texttt{other} in our figure. The \texttt{compute} is calculated according to the beginning of the first reconstruction task and the ending of the last, whereas the \texttt{other} column includes data transfer as well as the cloud service and resource allocation times.

The scaling efficiency of the \texttt{computing} component of the workflows is larger than 81\% relative to the one-node configuration. However, this changes when we consider \texttt{other} column. Since most of the operations, e.g, data transfer, (cloud) service calls and resource allocation, in \texttt{other} are not scalable, we see a drop in performance gain with the increasing number of nodes. Also, we set the granularity of the resource allocation request to one node, in other words, funcX requests only one node after exhausting all the available GPUs. This improves resource utilization while decreasing idle time, however it also introduces additional queue and resource initialization time during the execution. Overall, the \texttt{other} column contributes 3--37\% to the end-to-end execution time of the workflows, where we observe the worst case in eight-node configuration (mostly because of the shortest \texttt{compute} time.) The eight-node configuration still results in the shortest execution time with 3.9$\times$ speedup over one-node configuration. If we try to execute the same workflow using a local beamline workstation that has a single A100 Nvidia GPU, the total execution time would exceed 14 hours. Therefore, for large experimental datasets, e.g., hundreds of TBs, end-to-end execution times using local workstation can take weeks to finish. When we compare one-GPU configuration and eight-node configuration (64 GPUs), the speedup of using eight-node would be close to 29.8$\times$ (or under 30 mins.)
\section{Related Work}
\label{sec:related}
Ptychography has become one of the popular imaging techniques over the last decade~\cite{holler2014x,shapiro2017ptychographic,dong:2018:ptycho}. Several advanced reconstruction algorithms and parallelization techniques have been developed in order to address the computational requirements of ptychography workflows~\cite{yu2021ptycho,enfedaque2019high,marchesini2016sharp,nikitin2019photon,Nashed:14}. However, when coupled with complementary 3D imaging techniques, such as tomography or laminography, ptychography experiments can generate massive amounts of experimental data.
Many efficient and scalable 3D reconstruction techniques have been developed~\cite{xiao:2021:equilibrium,hidayetoglu2020petascale,bicer2015rapid}, but these techniques require high-performance computing resources that can only be accessed remotely, such as leadership computing facilities or large-scale user facilities.

Data analysis workflows at synchrotron light sources have been an active research area~\cite{klein2019interactive,Basham:fv5032,salim2019balsam,peterka2009configurable,bicer2017real}. 
CAMERA, an interdisciplinary project led by Lawrence Berkeley National Laboratory, investigates data analysis problems and workflows relevant to light sources~\cite{donatelli2015camera,ushizima2016ideal,Pandolfi:xe5032}. 
Similarly, Brookhaven National Laboratory initiated several programs that focus on NSLS-II facility and its workflows~\cite{allan2019bluesky,www:bnl-csi}. 
Jesse et al. from Oak Ridge National Laboratory uses a big data analysis framework to perform systematic analysis of ptychography~\cite{jesse2016big}.
These activities aim to provide efficient workflows and algorithms for analysis of imaging data at DOE facilities.
Scientific workflows have also been extensively studied in other areas by DOE, universities and other institutions~\cite{da2021workflows,nun:2020:wf,naughton2020software,turilli2019middleware,2013:wolstencroft:taverna,ossyra2019highly}.
Deelman et al. developed Pegasus workflow management system for transparent execution of scientific workflows, where workflows defined as directed acyclic graphs~\cite{2015:pegasus}. In this work, we implemented a workflow system that utilizes Globus services to enable execution of ptychographic reconstruction tasks on federated facilities and resources. Our system takes advantage of the Globus authentication infrastructure to manage identities~\cite{tuecke2016globus} and eases remote task execution and resource management~\cite{chard2020funcx,ananthakrishnan2018globus}.

Many machine learning (ML) techniques have been successfully used and integrated to light source and electron microscopy data analysis workflows to enhance and improve the quality of images and reconstructions~\cite{aslan2020distributed,kalinin2016big,venkatakrishnan2021algorithm}, including image denoising~\cite{wu2020deep,liu2020tomogan}, artifact reduction\cite{ziabari2020beam} and feature extraction~\cite{somnath2018feature}. 
These techniques can also be used for accelerating the performance of workflows and data acquisition~\cite{liu2019deep}. We plan to incorporate some of these advanced ML techniques in our workflow in the future.
\section{Conclusion}
\label{sec:conclusion}

We presented a system that unifies different facilities and resources to perform ptychographic data analysis. Our system establishes automated data analysis pipelines between (edge) instruments at synchrotron light sources and compute and storage resources in leadership computing facilities. Our system builds on several cloud-hosted services: funcX, a federated FaaS platform for remote execution of user-defined functions and resource management; Globus Flows, a workflow definition and execution service, for automation and definition of data analysis pipeline; and Globus, for efficient high performance and secure wide-area data transfers. We use high-performance ptychographic reconstruction software to maximize compute resource utilization.

We evaluated our system at APS and ALCF. Specifically, we simulated ptychographic data acquisition from a ptychography beamline at APS and reconstructed this data using the ThetaGPU cluster at ALCF. We observed significantly higher speedups and scalability efficiencies for large ptychography datasets compared to smaller ones. Since, our system utilizes Globus authentication service to integrate facilities, resources and services, data analysis pipelines can execute with little interference.  

\section*{Acknowledgments}

This material is based upon work supported by the U.S. Department of Energy, Office of Science, Basic Energy Sciences and Advanced Scientific Computing Research, under Contract DE-AC02-06CH11357. This research used resources of the Argonne Leadership Computing Facility and Advanced Photon Source, which are U.S. Department of Energy (DOE) Office of Science User Facilities operated for the DOE Office of Science by Argonne National Laboratory under the same contract. Authors acknowledge ASCR funded Braid project at Argonne National Laboratory for the workflow system research and development activities. Authors also acknowledge Junjing Deng, Yudong Yao, Yi Jiang, Jeffrey Klug, Nick Sirica, and Jeff Nguyen from Argonne National Laboratory for providing the experimental data. 
This work is partially supported by the Office of the Director of National Intelligence (ODNI), Intelligence Advanced Research Projects  Activity (IARPA), via contract D2019-1903270004.
The views and conclusions contained herein are those of the authors and should not be interpreted as necessarily representing the official policies or endorsements, either expressed or implied, of the ODNI, IARPA, or the U.S. Government.

%
%
\bibliographystyle{splncs04}
\bibliography{bibtex/ptycho}

\section*{Government License}
The submitted manuscript has been created by UChicago Argonne, LLC, Operator of Argonne National Laboratory (``Argonne''). Argonne, a U.S.\ Department of Energy Office of Science laboratory, is operated under Contract No.\ DE-AC02-06CH11357. The U.S.\ Government retains for itself, and others acting on its behalf, a paid-up nonexclusive, irrevocable worldwide license in said article to reproduce, prepare derivative works, distribute copies to the public, and perform publicly and display publicly, by or on behalf of the Government.  The Department of Energy will provide public access to these results of federally sponsored research in accordance with the DOE Public Access Plan. http://energy.gov/downloads/doe-public-access-plan.

\end{document}